\documentclass[twocolumn,prl]{revtex4}
\usepackage[hypertex]{hyperref}
\usepackage{amsmath,amssymb}
\usepackage{graphicx}
\usepackage{subfig}

\begin{document}

\title{Three-dimensional static vortex solitons in incommensurate magnetic crystals}
\author{A.B. Borisov}
\author{F.N. Rybakov}
\email{f.n.rybakov@gmail.com}
\affiliation{Institute of Metals Physics, Urals Branch of the Russian Academy of Sciences, Ekaterinburg, 620990,
Russia }
\date{Submitted December 29, 2009; revised March 4, 2010}

\begin{abstract}
A new type of three-dimensional magnetic soliton in easy-axis ferromagnets is predicted by taking
simultaneous account of the Dzyaloshinsky-Moriya interaction and an external magnetic
field. The numerically obtained static three-dimensional solitons with a finite energy are characterized
by a Hopf topological index $H=0$ and have a vortex structure. The structure of these solitons
and the dependence of their energy on the external field are determined. The asymptotic
behavior of these solitons is investigated and a necessary condition for their existence is
found.  
\end{abstract}

\maketitle

\section{I. Introduction}

At present a large number of magnetically ordered crystals
without an inversion center are known in which
exchange-relativistic interactions lead to the formation of
long-period magnetic structures whose periods are incommensurate
with the crystal-chemical periods \cite{bib:Izumov1,bib:Izumov2}. In the energy
for this system the exchange relativistic interaction is described
by terms which are linear in the first derivatives of
the magnetization (the Lifshitz invariant). The possibility
that they might have a key role in the formation of helicoidal
magnetic structures was first advanced in \cite{bib:DzHelicPart1}. Analytic
solutions \cite{bib:DzHelicPart3,bib:BarStef} showed that for certain values of the Dzyaloshinsky
constant modulated structures can exist in the system
with incommensurate periods. It was shown later \cite{bib:BorKis1,bib:BorKis2} that twodimensional
vortices can exist against the background of an
incommensurate structure in magnets with an ``easy-plane''
anisotropy according to a two-dimensional generalization of
the Dzyaloshinsky model (2D sine-Gordon). It has been
shown \cite{bib:BogdanovVortex1,bib:BogdanovVortex2} that two-dimensional magnetic vortices can exist in
magnetically ordered crystals with an ``easy axis'' anisotropy
for a certain range of external magnetic fields. Depending on
the parameters of the ferromagnetic medium and the external
magnetic field, both unstable and stable vortex states with
finite energies can develop \cite{bib:BogdanovVortex3,bib:BogdanovVortex4}. The structure of threedimensional
solitons and their domains of existence in these
magnetic materials have not yet been studied.

It is known that the Landau-Lifshitz equation without
dissipation,
\begin{equation}
\frac {\partial {\bf M}} {\partial t} = - \gamma \left[ {\bf M} \times {{\bf H_{eff}} }\right],\quad \label{eq:LLeq}
{{\bf H_{eff}} } = - \frac {\delta {E}} {\delta {\bf M}} ,\quad \gamma>0
\end{equation}
allows solutions in the form of three-dimensional precessional
solitons with a stationary profile under the following
conditions. If the energy $E$ of the ferromagnet includes only
the exchange energy,  
\begin{equation}
E _ {exch}=(\alpha / 2)\int \left({\partial _ {i}} {\bf M} \right)^2 d{\bf r}, \label{eq:ExchEnergy}
\end{equation}
i.e., the ferromagnet is isotropic, then \cite{bib:Cooper} nonstationary solitons
can exist which move uniformly along the axis of anisotropy
at a nonzero velocity. Here, however, the energy is proportional to the linear size of the soliton and the question
of stability with respect to collapse is especially acute. In
addition, it is not entirely right to neglect the energy of magnetic
dipole interactions, since demagnetizing fields can facilitate
the collapse of a soliton.

If, on the other hand, the energy $E$ of the ferromagnet
includes the uniaxial magnetic anisotropy energy,
\begin{equation}
E _ {anis}=(\beta / 2) \int \left({M _ x}^2+ {M _ y}^2\right) d{\bf r},\quad \beta>0 , \label{eq:AnizEnergy}
\end{equation}   
as well as the exchange energy $E _ {exch}$ then solitons with a
different structure can develop, and which either move
uniformly \cite{bib:Sut2001,bib:BorRyb2} or are stationary \cite{bib:IvKos1,bib:IvKos2,bib:DI,bib:BorRyb1}. The linear sizes of these
solitons are no longer determined by the value of an arbitrary
parameter, but depend on the characteristic magnetic length $l_0=\sqrt{\alpha / \beta}$. The contribution of the magnetic dipole interaction energy is really insignificant here when the magnetic medium has a high quality factor $Q=\beta/(4\pi)$.  

Nevertheless, in all the above cases we are speaking of
dynamic solitons, i.e., of the case where, when both the precession
frequency $\omega=0$ and the velocity $V=0$, a soliton cannot
exist. This has seriously inhibited experimental studies. A
real ferromagnetic medium is always dissipative and the lifetime
of a dynamic soliton may be very short. In order to slow
down or entirely stop the dissipation of solitons, it is necessary
to apply a special external actuation to the medium. In
addition, the necessary condition for observing solitons continues
to be the development of some mechanism for their
rapid formation, so that the average lifetime of a soliton exceeds
its formation time \cite{bib:IvKos2}. 

Static three-dimensional solitons cannot exist in any of
the above models of ferromagnets. This follows immediately
from the Hobart-Derrick theorem \cite{bib:Hobart,bib:Derrick}, which is also known
as the virial theorem \cite{bib:Raj}.

In many cubic ferromagnets without an inversion center,
the Dzyaloshinsky-Moriya energy makes a significant contribution
to their energy \cite{bib:DzRus,bib:Dz,bib:Mor}:
\begin{equation}
E _ {DM} = D \int {\bf M}\cdot\left(\nabla \times {\bf M}\right) d{\bf r}, \label{eq:DMEnergy}
\end{equation}
where $D$ is the Dzyaloshinsky constant. Then the total energy
of the ferromagnet is  
\begin{equation}
E = E_{exch} + E_{anis} + E_{DM} + E_{H}, \label{eq:FullEnergy}
\end{equation} 
where $E_{H}$ is the energy in an external magnetic field of
strength ${{\bf H}^{(e)}_0}=(0,0,H_0)$ relative to the ground magnetization
state ${\bf M}_{inf} = (0,0,M_0)$:
\begin{equation}
E _ {H} = H_0 \int \left(M_0 - M_z \right) d{\bf r}. \label{eq:HEnergy}
\end{equation} 

It turns out \cite{bib:Bogdanov1} that in the three-dimensional case the functional (\ref{eq:FullEnergy}) does not fall under the prohibition of the Hobart-Derrick theorem, so the existence of localized states remains
an open question. Here the necessary condition is $E_{DM}<0$ \cite{bib:Bogdanov2}. 

This paper is a study of static three-dimensional solitons
in the model of (\ref{eq:FullEnergy}). It is organized as follows: The structure
of the solitons, their stability, and the magnetic field
dependence of their energy are discussed in Section II. In
Section III the asymptotic behavior of the solutions is studied.
The resulting solitons are analyzed topologically in Section
IV. 

\section{II. THREE-DIMENSIONAL STATIC VORTEX SOLITONS}

It is reasonable to assume that three-dimensional solitons,
like plane vortices, can be in stable or unstable states.
The latter cannot be found by direct methods of unconditional
minimization of the energy functional. In order to
avoid this restriction, we shall seek a solution of the auxiliary
variational problem of minimizing the functional
\begin{equation}
F = E_{exch} + E_{anis} + E_{DM}, \label{eq:FuncEnergy}
\end{equation}        
with a fixed value of the integral
\begin{equation}
\mathcal{N} = \int (M _ 0 - {M _ z}) d{\bf r}.\label{eq:N2integral}
\end{equation}
We parametrize the magnetization vector ${\bf M}$ in terms of the
angular variables $\Theta$ and $\Phi$, as

\begin{equation}
{\bf M} = M _ 0 \cdot {\bf m} = M _ 0 \cdot (sin \Theta cos \Phi, sin \Theta sin \Phi, cos \Theta )
\label{eq:Mparam}
\end{equation}

In the presence of (\ref{eq:DMEnergy}), the energy density of the
ferromagnet is invariant with respect to simultaneous spatial
rotations by an arbitrary angle $\epsilon$ and the spatial coordinates,
and the magnetization vector, to
\begin{equation}
\varphi \rightarrow \varphi + \epsilon, \quad \Phi \rightarrow \Phi + \epsilon, \label{eq:Povorot}
\end{equation}
where $\varphi$ is the polar angle of a cylindrical coordinate system $(r,\varphi,z)$.
Thus, we shall study solitons with an axially symmetric
distribution of the polar angle of the magnetization
and a vortex structure for the azimuthal angle
\begin{equation}
\Theta=\theta(r,z), \quad \Phi=\varphi + \phi(r,z) \label{eq:Q1vortex}
\end{equation}
with the boundary conditions
\begin{equation}
\theta\rightarrow0 \quad (|{\bf r}|\rightarrow \infty)
\end{equation}
at infinity and the symmetries
\begin{equation}
\theta(r,-z)=\theta(r,z), \quad \phi(r,-z)=\pi - \phi(r,z).
\end{equation}
No boundary conditions are imposed initially at the axis of
symmetry but show up as a result of the numerical procedure.

The functions $\Phi$ and $\Theta$ are solutions of the posed variation
problem and, at the same time, extremal functionals of
the energy (\ref{eq:FullEnergy}) for a certain value of $H_0$. The problem actually
involves finding the functions $\phi(r,z)$ and $\theta(r,z)$ that specify
the unit vector field on the half plane
\begin{equation}
{\bf n}(r, z) = (sin \theta cos \phi, sin \theta sin \phi, cos \theta ), \label{eq:nParam}
\end{equation}
To calculate the value of $H_0$ corresponding to extremum of
the energy functional, we use the necessary condition for an
extremum: 
\begin{equation}
\frac{d}{d\lambda} E({\bf m}({\bf r} + \lambda \cdot {\bf r})) \Bigr|_{\lambda=0} = 0.
\end{equation}
from which we immediately obtain
\begin{equation}
H_0 = \frac{ - 2 E_{DM} - E _ {exch} - 3 E _ {anis} }{3 \mathcal{N} }. \label{eq:HfromIntegrals}
\end{equation}

The variational problem was solved by the same numerical
method as for conditional minimization in \cite{bib:BorRyb1}. The initial
configuration for the field was specified by the same functions
as in \cite{bib:BorRyb2} for determining the structure of moving
magnon droplets. To test the resulting configurations, we calculated
the corresponding external magnetic field strength $H_0$
according to (\ref{eq:HfromIntegrals}), and, at the grid points, the values of
the parametrizing angles $\phi(r,z)$ and $\theta(r,z)$ and their first
and second order spatial derivatives. After this, the discrepancy
in the Euler-Lagrange equations for the energy functional
(\ref{eq:FullEnergy}) was calculated directly: 
\begin{multline}
- \frac{sin 2\theta} {2} \left( \frac{1}{r^2}+\frac{1}{{l_0}^2} + ({\partial _ {r}}{\phi})^2 + ({\partial _ {z}}{\phi})^2 \right) +{} \\
{}+ \frac{4 \chi}{\pi l_0}\left(\frac{sin 2\theta} {2} {\partial _ {z}}{\phi} + (sin\theta)^2 (cos\phi{\partial _ {r}}{\phi} + \frac{sin\phi}{r})\right)+{} \\
{}+ \Delta_2\theta - (h/{{l_0}^2})sin\theta = 0,\label{eq:EL1}
\end{multline}
\begin{multline}
-(4 \chi / \pi l_0)\left(cos\theta {\partial _ {z}}{\theta} + cos\phi sin\theta {\partial _ {r}}{\theta}\right) +{} \\
{}+ 2 cos\theta \left({\partial _ {r}}{\theta}{\partial _ {r}}{\phi} + {\partial _ {z}}{\theta} {\partial _ {z}}{\phi} \right) + sin\theta \Delta_2\phi=0, \label{eq:EL2}
\end{multline}
where the differential operator
\begin{equation}
\Delta_2=\frac{1}{r}\frac{\partial}{\partial r} + \frac{{\partial^2 }}{{\partial r}^2} + \frac{{\partial^2 }}{{\partial z}^2}
\end{equation}
and the dimensionless parameters                                                                                                             
\begin{equation}
\chi=\frac{\pi D}{2 \sqrt{\alpha\beta}},\quad h=\frac{H_0}{\beta M_0},
\end{equation}
as in \cite{bib:BogdanovVortex4}.

\begin{figure}
\includegraphics[width=1.0\columnwidth]{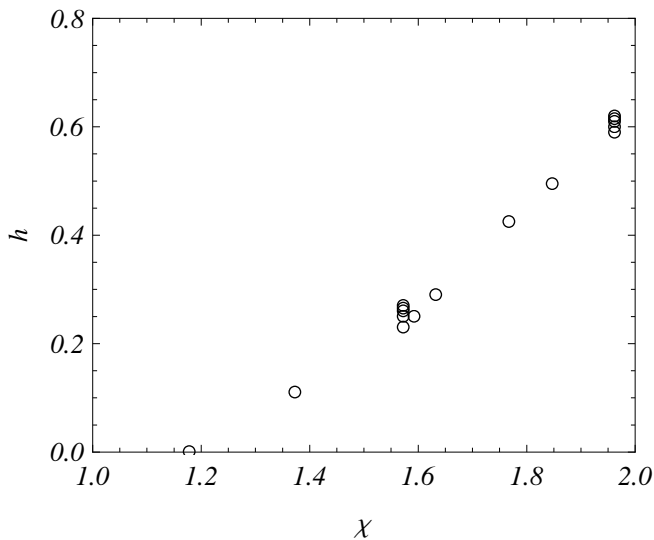}
\caption{A $\chi - h$ diagram of the results of the numerical calculations. Each
point corresponds to a successful computational cycle, i.e., to establishing
the existence of a soliton and determining its structure.}
\label{F:fig1}
\end{figure}

\begin{figure}
\subfloat[]{\label{F:fig2a}\includegraphics[width=1.0\columnwidth]{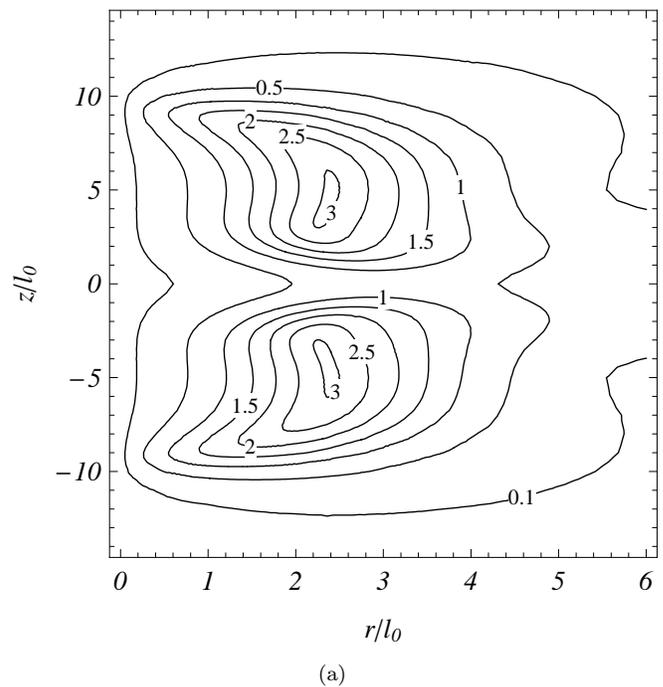}}\\
\subfloat[]{\label{F:fig2b}\includegraphics[width=1.0\columnwidth]{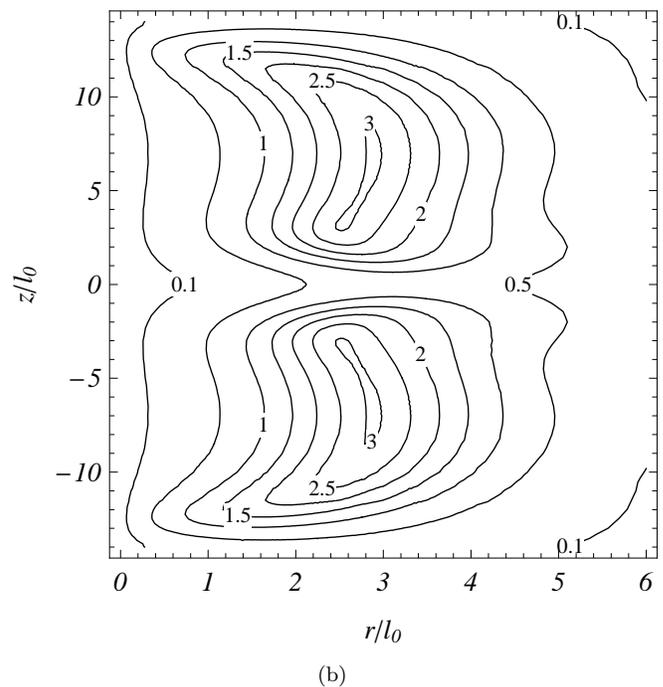}}
\caption{Contours of the angle $\theta$ of a soliton configuration for two values of
the parameter $h$ with constant $\chi=1.96$: $h=0.60$ (a), $h=0.62$ (b).}
\label{F:fig2}
\end{figure}

In calculations using the method described here it is necessary
every time to specify the value of the integral (\ref{eq:N2integral}) and
the parameter $\chi$. And only if the calculation leads to a positive
result (to a soliton solution) can be calculate a corresponding
value of the parameter $h$. In addition, if minimizing
the functional does not yield a soliton solution, this does not mean the latter is nonexistent. This sort of situation can arise,
for example, when the modelling region is not the right size
(the soliton is too big). In light of this, only points corresponding
to positive results are plotted in Fig.\ref{F:fig1}.   

\begin{figure}
\includegraphics[width=1.0\columnwidth]{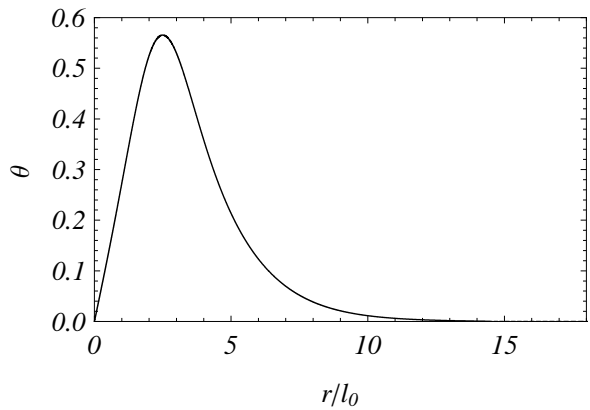}
\caption{$\theta$ as a function of the coordinate $r$ at $z=0$ for a soliton with the
parameters of a ferromagnetic medium $\chi=1.57$, $h=0.25$.}
\label{F:fig3}
\end{figure}

\begin{figure}
\subfloat[]{\label{F:fig4a}\includegraphics[width=0.5\columnwidth]{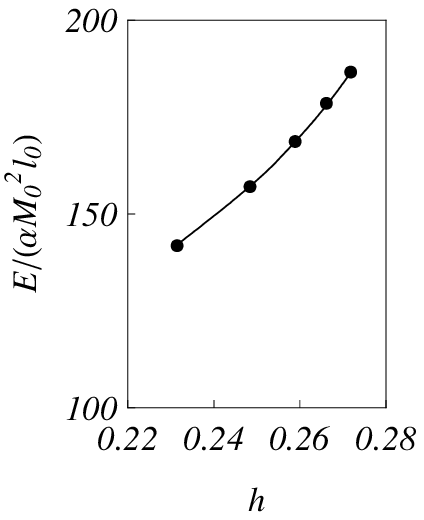}}
\subfloat[]{\label{F:fig4b}\includegraphics[width=0.5\columnwidth]{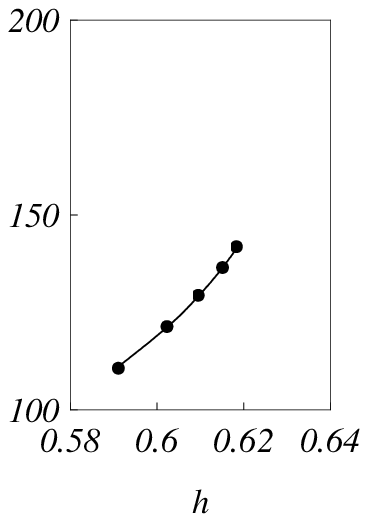}}
\caption{Soliton energy as a function of the parameter $h$ for two values of the
parameter $\chi$ for a ferromagnetic medium: 1.571 (a) and 1.963 (b). The
points indicate the results of numerical calculations.}
\label{F:fig4}
\end{figure}

Figure \ref{F:fig2} shows contours of constant angle $\theta$ for a typical
soliton in the $(r,z)$ half plane for two values of the external
magnetic field. In a three-dimensional space a constant value
of the angle $\Theta$ corresponds to one or several toroids with a
complicated envelope. 

The profile $\theta(r,z)$ of a soliton for $z=0$ shown in Fig.\ref{F:fig3}
shows that the characteristic size of the localized structures
being studied here is on the order of $10 l_0$.

Figure \ref{F:fig4} is a plot of the energy of the soliton as a function
of the external magnetic field. As the magnetic field is
increased, the energy also rises. As these graphs show, to
excite a soliton with a given energy in a material with a
higher value of $D$, it is necessary to increase the external
magnetic field strength $H_0$.

It turned out that all the solitons that were found are
stable with respect to scaling perturbations:
\begin{equation}
\dfrac{d^2}{d\lambda^2} E({\bf m}({\bf r} + \lambda \, {\bf r})) \Bigr|_{\lambda=0} > 0,
\end{equation}
\begin{equation}
\dfrac{d^2}{d\lambda^2} E({\bf m}(x+\lambda \, x,y,z)) \Bigr|_{\lambda=0} > 0,
\end{equation}
\begin{equation}
\dfrac{d^2}{d\lambda^2} E({\bf m}(r,\varphi,z+\lambda \, z)) \Bigr|_{\lambda=0} > 0,
\end{equation}
\begin{equation}
\dfrac{d^2}{d\lambda^2} E({\bf m}(r+\lambda \, r,\varphi,z)) \Bigr|_{\lambda=0} > 0.
\end{equation}
Nevertheless, the extrema that were found do not provide a
minimum for the energy functional (\ref{eq:FullEnergy}) in a strict mathematical sense.
We examined the stability of the solitons with
respect to small perturbations. A special type of functions $\delta\phi(r,z)$ and $\delta\theta(r,z)$ (complex perturbations) was found such that
\begin{equation}
\dfrac{d^2}{d\lambda^2} E( \phi +  \lambda \, \delta\phi, \theta + \lambda \, \delta\theta) \Bigr|_{\lambda=0} < 0.
\end{equation}
However, even if the structures being studied here turn out to
be unstable, they may have a rather long lifetime.

\section{III. ASYMPTOTIC BEHAVIOR}

Data from the numerical calculations show that with distance
from the center of a soliton the azimuthal angle of the
vector ${\bf n}$ (\ref{eq:nParam}) becomes a linear function of the coordinate $z$, i.e. $\phi(r,z)\rightarrow 2\pi m + \pi/2 + k z$ as $|{\bf r}|\rightarrow\infty$, $m\in{\mathbb{Z}}$. In addition,
the soliton structure found as a result of minimization shows
that $\theta(0,z)=0$.  Since the angle $\theta$ goes monotonically to zero
as $|{\bf r}|\rightarrow\infty$, the initial system of Eqs. (\ref{eq:EL1}) and (\ref{eq:EL2}) can be
linearized for small $\theta$ with larger values of $|{\bf r}|$:
\begin{equation}
-\left(k^2 - \frac{4 \chi k}{\pi l_0} + \frac{1+h}{{l_0}^2} + \frac{1}{r^2}\right)\theta + \Delta_2 \theta=0
\label{eq:AsympThetaEq}
\end{equation} 
\begin{equation}
(-(4 \chi / \pi l_0) + 2k){\partial _ {z}}{\theta}=0.
\label{eq:AsympPhiEq}
\end{equation} 
It follows at once from (\ref{eq:AsympPhiEq}) that the constant k is determined
by the Dzyaloshinsky constant, as
\begin{equation}
k = 2 \chi / (\pi l_0) = D/\alpha.\label{eq:k} 
\end{equation}
Using the auxiliary function
\begin{equation}
\psi(r,z,\varphi)=\theta(r,z) cos(\varphi)
\label{eq:Psi}
\end{equation}
it is convenient to rewrite Eq. (\ref{eq:AsympThetaEq}) in the form
\begin{equation}
\Delta\psi=\gamma^2 \psi,
\label{eq:Helmholtz}
\end{equation}
where $\Delta$ is the laplacian operator in three dimensional space and 
\begin{equation}
\gamma=\sqrt{1+h-(4\chi^2/\pi^2)  }/{l_0}.\label{eq:gamma}
\end{equation}

We shall seek a solution for (\ref{eq:Psi}) and (\ref{eq:Helmholtz}) subject to
the conditions
\begin{equation}
\begin{cases}
\theta\geqslant0, r\geqslant0,
\\
\theta(0,z) = 0,
\\
\theta\rightarrow 0 \quad (|{\bf r}|\rightarrow\infty),
\\
\theta(r,-z)=\theta(r,z).
\end{cases}
\label{eq:GranUsl}
\end{equation}

Note that in the special case of $\gamma=0$, Eq.(\ref{eq:Helmholtz}) transforms
to the Laplace equation
\begin{equation}
\Delta\psi=0,
\label{eq:Lapl}
\end{equation}
and has an interesting electrostatic analog: the function $\psi$ is
the potential for the electric field of a system consisting of
charges distributed along a circle of arbitrary, but finite, radius
$r=a$, at $z=0$, with a linear density proportional to $cos(\varphi)$
and an infinitely long, grounded filament passing along the
axis of symmetry, where the potential of the filament, like the
potential at infinity, is zero. In fact, Eq.(\ref{eq:Lapl}) permits separation
of variables in toroidal coordinates $(u,v,\varphi)$,  
\begin{align}
tanh(u) &= (2 a r)/(a^2 + r^2 + z^2),\quad 0\leqslant u < \infty, \\
cot(v) &= (a^2 - r^2 - z^2)/(2 a z),\quad 0\leqslant v < 2\pi,
\end{align}
and has the exact solution \cite{bib:Andrews}: 
\begin{equation}
\psi = C \sqrt{a/r} Q_{1/2}(\chi) cos(\varphi), \\
\label{eq:LaplSol}
\end{equation}
where $C$ is an arbitrary positive constant, $\chi=coth(u)$, $Q_{1/2}(\chi)$ is the Legendre function of the second kind of order
$1/2$. This function can be written as the integral \cite{bib:Lebedev}:
\begin{multline}
Q_{1/2}(\chi) =\int_{0}^{\pi} {\frac {cos\xi}{\sqrt{2\chi-2cos\xi}} } {d\xi} = {} \\ 
{} = \chi k_{\chi} K(k_{\chi}) - \frac{2}{k_{\chi}} E(k_{\chi}),\quad  k_{\chi}=\sqrt{\frac{2}{1+\chi}},
\end{multline}
where $K$ and $E$ are the complete elliptical integrals of the
first and second kind, respectively. The solution (\ref{eq:LaplSol}), like
every linear combination of similar solutions, satisfies Eq.
(\ref{eq:GranUsl}) and yields the following asymptotic representation for the function
\begin{equation}    
\theta(r,z) \sim (r^2+z^2)^{-3/2} r \quad (|{\bf r}|\rightarrow\infty), \quad \gamma=0. 
\label{eq:TeLaplAsympt}
\end{equation}

The result (\ref{eq:TeLaplAsympt}) can be derived formally in another way. If $\vartheta$ is the polar angle of the spherical coordinate system $(\rho,\vartheta,\varphi)$ with $\rho = |{\bf r}| = \sqrt{r^2+z^2}$, then separating the variables in (\ref{eq:Lapl}) and using (\ref{eq:Psi}) yields a basis of the functions of the form
\begin{equation}
\psi_n=C \rho^{-n-1} P_{n}^1(cos\vartheta)cos(\varphi).
\end{equation}

Given the imposed conditions (\ref{eq:GranUsl}), $n$ is a positive odd
number and in the asymptotic limit, only $n=1$ ``survives''. This immediately yields the asymptote (\ref{eq:TeLaplAsympt}).  

If $\gamma \neq 0$, then the Helmholtz equation (\ref{eq:Helmholtz}) is solved by
separation of variables in spherical coordinates. Then, given
(\ref{eq:Psi}), we find the basis of the functions in the form      
\begin{equation}
\psi_n=\frac{C}{\sqrt{\rho}} K_{n+1/2}(\gamma \rho) P_{n}^1(cos\vartheta)cos(\varphi),
\end{equation}
where $K_{\nu}(\chi)$ is the modified Bessel function of the second
kind. As a result, we obtain the final formula for the
asymptotic behavior:
\begin{equation}
\theta(r,z) \sim \frac{(1+\gamma \sqrt{r^2+z^2}) \, r}{e^{\gamma \sqrt{r^2+z^2}}(r^2+z^2)^{3/2}} \quad (|{\bf r}|\rightarrow\infty).
\end{equation}
This implies the necessary condition for the existence of a
three-dimensional vortex soliton: $\gamma>0$. Then, according to
(\ref{eq:gamma}),
\begin{equation} 
1+h-(4\chi^2/\pi^2)>0.\label{eq:KriteriiHomo}
\end{equation}
This result is in qualitative agreement with the plot in Fig.\ref{F:fig1}.
Note that this condition is the instability criterion for a helicoidal
structure in uniaxial magnets \cite{bib:BarStef}.

When the condition
\begin{equation}
1+h-(4\chi^2/\pi^2)<0,\label{eq:KriteriiHelic}
\end{equation}
holds, the ground state of the hamiltonian (\ref{eq:FullEnergy}) corresponds to
the helicoidal structure
\begin{equation}
cos(\Theta)=\frac{h}{(4\chi^2/\pi^2) - 1}, \quad \Phi=k z, \label{eq:Helicoid}
\end{equation}
where the step size of the spiral $\lambda=2\pi/k$ and $k$ is the same
as in (\ref{eq:k}). This sort of structure can be classified as a
ferromagnetic spiral (FS) \cite{bib:Izumov1,bib:Izumov2}. Here the energy density
\begin{equation}
w=-\frac{{M_0}^2 \beta}{2 \pi^2} \frac{(1+h-(4\chi^2/\pi^2))^2}{(4\chi^2/\pi^2)-1}.
\end{equation} 
This implies that the condition (\ref{eq:KriteriiHomo}) holds, a uniform stable
state $\theta=0$ is realized in the system. Thus, the solitons we
have been studying are localized excitations of a uniform
state.

The volume energy density of the soliton, $w$, falls off
exponentially with distance from the center of the soliton, i.e.,
\begin{equation}
w(\rho,\vartheta) \sim - \frac{cos(k \rho \, cos\vartheta)(sin \vartheta)^2 }{e^{\gamma \rho}\rho} \quad (\rho\rightarrow\infty).
\end{equation}  

\section{IV. THE TOPOLOGY OF VORTEX SOLITONS}

\begin{figure}
\includegraphics[width=1.0\columnwidth]{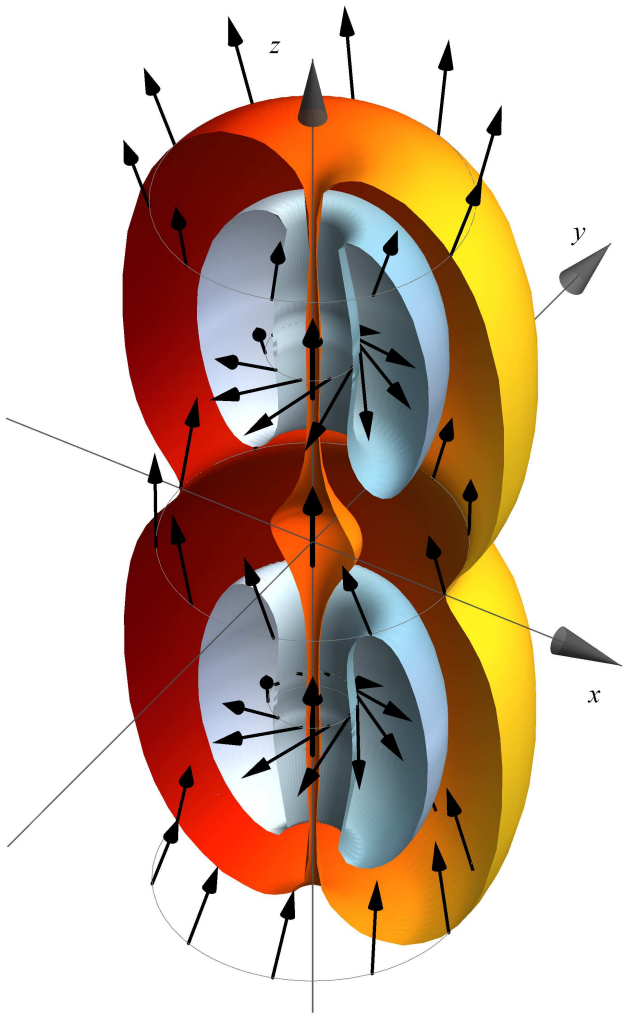}
\caption{The distribution of the magnetization for a soliton with $h=0.11$, $\chi=1.37$. The toroids constructed in the interval $0<\varphi<\pi$ correspond to two values of the angle $\Theta$: for the outer toroid $\Theta=0.3$ and for the inner, $\Theta=2.0$}
\label{F:fig5}
\end{figure}

The unit vector field ${\bf m}$ is continuous and specified at
each point of ${\mathbb R}^3$ space, while in the asymptotic limit ${\bf m}\rightarrow(0,0,1)$ as $|{\bf r}|\to\infty$. These kinds of fields map the space ${\mathbb R}^3\cup \{\infty \}$ onto a two-dimensional sphere ${\mathbb S}^2$ are classified as
belonging to homotopic class $\pi_3({\mathbb S}^2)={\mathbb Z}$, and are characterized
by an integral Hopf index $H$:
\begin{equation}
H=-\frac{1}{(8\pi)^2}\int{\bf F}\cdot{\bf A}\,d{\bf r},\label{eq:Hdefault}
\end{equation}
where $F_{i}=\epsilon_{ijk}{\bf m}\cdot(\nabla_j{\bf m}\times \nabla_k{\bf m})$ and $(\nabla \times {\bf A}) = 2 {\bf F}$.
The expression for the Hopf index $H$ of the fields (\ref{eq:Q1vortex}) simplifies \cite{bib:KUR,bib:Glad} to
\begin{equation}
H = \frac {1} {4 \pi}\int_{-\infty}^{\infty} {\int_{0}^{\infty} {{\bf n}\cdot\left[{\partial}_r {\bf n} \times {\partial}_z {\bf n} \right]} d{r}}d{z}.\label{eq:HNumber}
\end{equation}

If the index $H$ in (\ref{eq:HNumber}) is nonzero, then the soliton can
be classified as a toroidal hopfion \cite{bib:Kamchatnov,bib:KBK,bib:Fadd2,bib:KUR,bib:Glad,bib:BorRyb2}. A numerical calculation
shows that for all the solitons obtained here, $H = 0$. An
example of a nontopological soliton with $H=0$ is a vortexfree
magnon droplet \cite{bib:IvKos1}.
For the class of objects studied here,
as in Fig.\ref{F:fig5}, one can see a structure of vortex wheels. The
solitons found here can be classified as topological in the
sense that they form an ensemble of ``hopfion-antihopfion'' pairs.

\section{V. CONCLUSION}
It has been shown here that, within a definite region of
parameters, three-dimensional vortex solitons can exist in incommensurate
magnetically ordered crystals. The structure
of topological solitons with a nonzero Hopf index $H$ (hopfions) will be
described in later papers. The static solitons obtained here do
not have the disadvantages of precessing magnon droplets \cite{bib:IvKos2}.
Thus, despite their possible instability, they may have a sufficiently
long lifetime that they can be observed in real experiments.
In particular, if the external magnetic field is
changed suddenly to a state (\ref{eq:KriteriiHomo}) for a stable helicoidal structure
(\ref{eq:Helicoid}), then the magnet will end up in an unstable phase
and the transition to a uniform state may be accompanied by
the appearance of solitons. It is also possible that within
some range of the parameters $\chi-h$ that has not been studied in
this paper, solitons of this type will be stable from a mathematical
standpoint.

We thank B. A. Ivanov for the invitation to submit this
article for the issue of this journal devoted to the 80-th birthday
of V. G. Bar'yakhtar, whose scientific activity and school
have always made significant contributions to the development
of the physics of magnetic phenomena.

\end{document}